# Numerical Solution of Cylindrically Converging Shock Waves

## M. Shabouei, R. Ebrahimi and K. Mazaheri Body

***Abstract***—The cylindrically converging shock wave was numerically simulated by solving the Euler equations in cylindrical coordinates with TVD scheme and MUSCL [1] approach, using Roe's approximate Riemann solver [2] and super-bee nonlinear limiter [3]. The present study used the in house code developed for this purpose. The behavior of the solution in the vicinity of axis is investigated and the results of the numerical solution are compared with the computed data given by Payne [4], Lapidus [5], Abarbanel and Goldberg [6] Sod [7] and Leutioff et al. [8].

## I. INTRODUCTION

The cylindrical converging shock problem was computed by Payne [4], Lapidus [5], Abarbanel and Goldberg [6], Satofuka [9], Sod [7], Ben-Artzi and Falcovitz [10], Leutioff *et al.* [8] and Matsuo *et al.* [11]. The finite difference method (the Lax scheme) was first applied to the cylindrical shock tube problem by Payne. Lapidus used the "Cartesian method" to compute flows with radial symmetry. Comparisons with Payne's solutions indicated that there was a difference in the shock speeds. Abarbanel and Goldberg computed the same problem using an iterative procedure based on the modified Lax-Wendroff method. The results were compared with Payne's solutions and it differed markedly from this solutions. Satofuka used other difference schemes to solve the present imploding shock problem. Sod solved this problem and captured sharp discontinuity throughout the calculation by combining the Glimm's random choice method [12] and time splitting. The results were compared with the finite difference methods and it indicated a difference in the time at which the shock reached the axis. Ben-Artzi and Falcovitz proposed "GRP" (generalized Riemann problem) method as an upwind second-order scheme. Cylindrical shock tube problems were tested and it was in very good agreement with those of Abarbanel and Goldberg and Sod. Leutioff *et al.* [8] used finite difference scheme of invariant character proposed by Rusanov [13].The results of the numerical calculation are compared with Abarbanel and Goldberg and Payne. Differences between the results are due to the non-invariant character of the other methods. Matsuo *et al.* [11] simulated propagation of the converging shock wave, which is assumed to be generated by an instantaneous energy release on a rigid cylindrical wall. The results compared whit selfsimilar solution, the random choice method, the method of characteristics and the second-order finite difference method with artificial viscosities. They all agreed with one another except for the focusing stage.

A theoretical investigation of converging shock waves by means of similarity solution in a perfect gas has been presented by Guderley [14], Butler [15], Zeldovich and Raizer [16], Whitham [17] and Lazarus [18]. Van Dyke and Guttmann [19] and Hafner [20] also proposed alternative methods to solve the convergent shock problem. More recently Sharma and Radha [21], Bilbao and Gratton [22], Samtaney and Pullin [23], Hidalgoa and Mendozab [24] have derived other similarity solution to this problem. A famous self-similar solution that describes implosions is the convergent shock wave solved first by Guderley. Guderley showed that the strength of the shock wave varies inversely in case of cylindrical or spherical symmetry. A review on theoretical solution has been written by Meyer-ter-Vehn and Schalk [25], where additional references can be found.

Over the years many experimental works have been done on cylindrically converging shock waves. Perry and Kantrowitz [26], Wu *et al.* [27] , [28], Takayama *et al.* [29], Watanabe *et al.* [30] and Eliasson *et al.* [31] , [32] investigated such experiments. Perry and

Mohammad Shabouei is MSc. Student at Dept. of Aerospace Eng., Khaj-Nasir Toosi Univ. of Tech., Tehran, Iran, (corresponding author to provide e-mail: shabouei@sina.kntu.ac.ir, m.shabouei@gmail.com).

Reza Ebrahimi is Associate Professor at Dept. of Aerospace Eng., Khaj-Nasir Toosi Univ. of Tech., Tehran, Iran, (e-mail: REbrahimi@kntu.ac.ir).

Kiumars Mazaheri Body is Associate Professor at Dept. of Mech. Eng., Tarbiat Modares Univ., Tehran, Iran (e-mail: kiumars@modares.ac.ir).

Kantrowitz and Wu *et al.* studied the production and stability of convergent cylindrical shock waves. Takayama *et al.* also investigated the stability of converging cylindrical shock waves in air. Watanabe *et al.* studied Shock wave focusing in a vertical annular shock tube and the light emission from a converging shock wave was investigated experimentally by Eliasson *et al.* at their two reported works.

The objective of the present work is to numerically study a cylindrically converging shock wave and use the in house code developed for this purpose in order to identify fluid characteristics at various initial conditions.

It is common to use annular shock tubes to create and study converging shock waves. The converging shocks are often visualized by either schlieren photographs or interferograms taken during the focusing process. These methods give a measure of the shock position and shape development as a function of time. With these techniques it is not possible to measure other quantities, like temperatures and pressures [32]. But with annular shock tubes can perfectly compute flow properties. So, in the present we use a cylindrical shock tube with a cylindrical diaphragm of radius $r_0$ separates two uniform regions of a gas at rest where the outer pressure and density being higher than the inner ones. At time $t = 0$ the diaphragm is ruptured and the inward flow starts. A shock wave is created and travels into the low pressure region followed by a contact discontinuity towards the axis, while an expansion fan travels outward into the high pressure region.

In this system of differential equations the axis is a singular point of the solution due to the polar coordinates. To remove the singularity nature at the centre, Payne [4] used extrapolation as a trick and Lapidus [5] proposed the Cartesian method which employed the equations in Cartesian coordinates.

In order to verify the CFD scheme, the cylindrically converging shock wave and the shock tube problem were first calculated and compared with the data found in literatures and exact solution, respectively. The CFD scheme is then used to study the cylindrically converging shock. Variations of the pressure and temperature profiles are traced step by step during the simulations, and their dependence on the initial conditions is discussed.

Following our previous study [33], the present work also involves a distinct point: the initial pressure and density ratio is much larger than that in the preceding studies. Actually it ranges from tens to thousands.

## II. Governing Equations and Numerical Simulations

In this paper we consider the numerical study of the compressible one-dimensional multi-component Euler equations for an inviscid, non-heat conducting, and cylindrically symmetric flow describing the unsteady motion of converging cylindrical shock waves. The equations can be written in conservation form for an ideal gas in cylindrical coordinates as following

$$\frac{\partial (A\,U)}{\partial t} + \frac{\partial (A\,F)}{\partial r} = G'  \qquad (1)$$

where

$$U = \left\{ \begin{array}{c} \rho \\ \rho u \\ \rho E \end{array} \right\}, \quad F(U) = \left\{ \begin{array}{c} \rho u \\ \rho u^2 + P \\ u\,(\rho E + P) \end{array} \right\},$$

$$G'(U) = \left\{ \begin{array}{c} 0 \\ P\,\dfrac{dA}{dr} \\ 0 \end{array} \right\}$$

whereas $\rho$ is the density, $u$ is the velocity, $P$ is the pressure, $E$ is the total energy; $t$ and $r$ are the independent variables which represent time and the space coordinate of symmetry respectively and $A$ is the flow cross section. Here

$$E = \frac{P}{\rho\,(\gamma - 1.0)} + \frac{1}{2} u^2$$

where $\gamma$ is the ratio of specific heats and in this work a constant greater than 1.0. After simplicity, the system (1) can be written in the form

$$\frac{\partial U}{\partial t} + \frac{\partial F}{\partial r} = G  \qquad (2)$$

where

$$G = -\frac{1}{A}\frac{dA}{dr} \left\{ \begin{array}{c} \rho u \\ \rho u^2 \\ u(\rho E + P) \end{array} \right\}$$

and for cylindrical symmetry

$$G = -\frac{1}{r}\left\{\begin{array}{c} \rho u \\ \rho u^2 \\ u\left(\rho E + P\right)\end{array}\right\}$$

The method of solution consists of a combination of TVD scheme and operator splitting. To use method of splitting, first remove the inhomogeneous geometry source term $G(U)$ from the system (2). Thus we will solve the system

$$\frac{\partial U}{\partial t} + \frac{\partial F}{\partial r} = 0 \qquad (3)$$

which reduces to one-dimensional Euler equations in Cartesian coordinates. The system (3) is integrated using the explicit Lax-Wendroff scheme. To obtain a sharp and monotone discontinuities representation, a TVD scheme and MUSCL [1] approach based on Roe's approximate Riemann solver [2] is employed, along with the super-bee limiter [3]. This scheme is second-order accurate both in space and time.

Once system (3) has been solved, the system of ordinary differential equations

$$\frac{dU}{dt} = G \qquad (4)$$

is solved, the solution of system (3) being used to obtain the geometry source term $G(U)$ in (4). In the present work, we used the 2nd order Runge-Kutta method to solve system (4).

## III. DIMENSIONLESS EQUATIONS

The dependent variables are non-dimensionalized with respect to the undisturbed fluid properties. Density is dimensionless with $\rho_0$ and pressure with $\gamma P_0$ where $\gamma = 1.4$. For the velocity, the sound speed of the undisturbed fluid, $c_0$, is used as the reference. The length scale is $r_0$, the radius of the diaphragm. Hence, the length scale does not change for different gases. The time is dimensionless with the length scale divided by the sound speed in the undisturbed fluid (i.e., $t_0 = r_0/c_0$). The gas is assumed to behave like an ideal gas which has the non-dimensional equation of state in the form $P = \rho T$.

## IV. CODE VERIFICATION

Three numerical tests have been conducted to validate the present simulation. The first two tests are some extreme cases of one-dimensional unsteady shock tube problems that were compared with the exact solution and with numerical results of lattice Boltzmann method (LBM) [34] to validate the TVD scheme simulation. Generally speaking, the computed properties (pressure, density, velocity and Mach No.) profiles are in very good agreement with the exact and LBM solutions of shock tube for wide range of initial conditions.

As the third case, cylindrically converging shock wave data are compiled from the literature and compared to the present work to establish validation. In general the overall trend of the computed results agrees with those of previous simulations such as Payne [4], Lapidus [5], Abarbanel and Goldberg [6], Sod [7] and Leutioff *et al.* [8]. However, there exist some discrepancies. The major difference is the time at which the shock reaches the axis. In all the methods used earlier the shock converges at the axis of symmetry at a later or slightly earlier time than in our case. Considering different assumptions in the numerical simulations and the theoretical analysis, the differences are understandable which will be discussed in next section.

In conclusion, the code is well verified and the numerical solutions are acceptable with reasonable accuracy.

## V. RESULTS AND DISCUSSION

For the present numerical simulation the initial pressure and density ratio as initial conditions, are varied in a wide range from tens to thousands. In the profiles given below initial pressure and density ratio have the same value and the temperature in the two initial regions inside and outside the diaphragm is assumed to be equal. With increasing time the temperature field becomes non-uniform. We took $r_0$, the radius of the diaphragm 1.0 and the mesh size $\Delta r = 0.005$. The time step $\Delta t$ is chosen such that CFL condition, $\nu$, is satisfied, i.e.

$$\max(|u| + c)\Delta t / \Delta r \leq \nu$$

where $\nu$, in the present work is set to 0.5.

In the following, results cover the relatively wide ranges of initial conditions from weak shock (the ratio of initial conditions 4), intermediate strength (the ratio 10) to strong shock (the ratio 20).

In figure 1, initial pressure and density ratio is 4. The pressure and temperature distribution is displayed at time intervals of 0.1 and 0.2 respectively. The shock appears as a rapid variation in $P$ and $T$ which is completely sharp. As time increases and the shock travels towards the axis and the strength of it increases.

contact discontinuity (similar to the shock wave) is completely sharp.

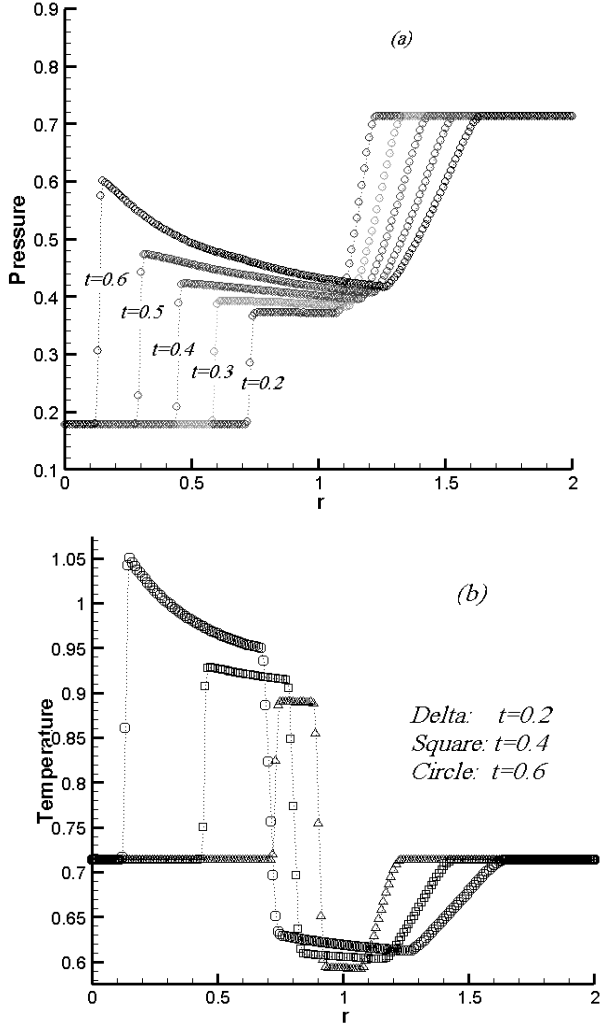

Fig. 1 Pressure (a) and Temperature (b) profiles at several time steps for initial pressure and density ratio 4.

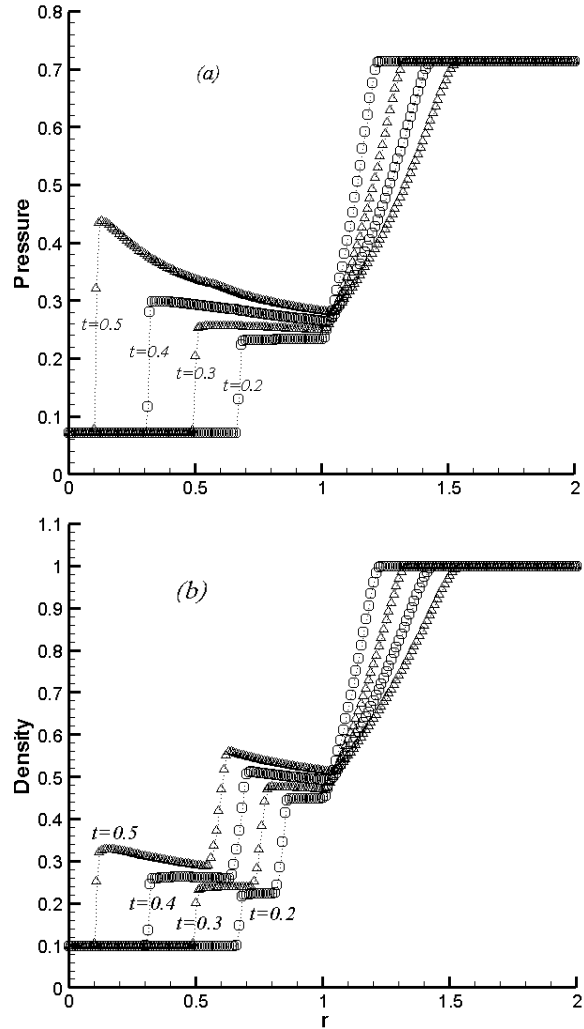

Fig. 2 Pressure (a) and Density (b) profiles at time intervals of 0.1 for initial pressure and density ratio 10.

After the passage of the shock, the pressure at the points behind it continues to rise. When the shock reaches the axis, maximum but finite value of pressure in convergence is occurred.

In the temperature profile, the basic properties of shock are similar to those of the pressure distribution, but the rise in temperature across the shock is higher. In addition to the shock wave, in the temperature graph a contact discontinuity appears. The contact surface travels towards the axis behind the converging shock. As a result of using the super-bee limiter, the

In figure 2, initial pressure and density ratio is 10. The pressure and density profile is shown at time intervals of 0.1. In the density profile, the basic properties of shock are similar to those of the pressure and temperature distribution, except that the rise in density across the shock is smaller, corresponding to a temperature increase. In the density profile such as temperature, a contact discontinuity appears.

In figure 3, initial pressure and density ratio is rise to 20. The pressure and temperature distribution is displayed at time intervals of 0.1. As predicted by Payne [4] the stronger shock increases in strength more rapidly than the weak shock as it approaches the axis.

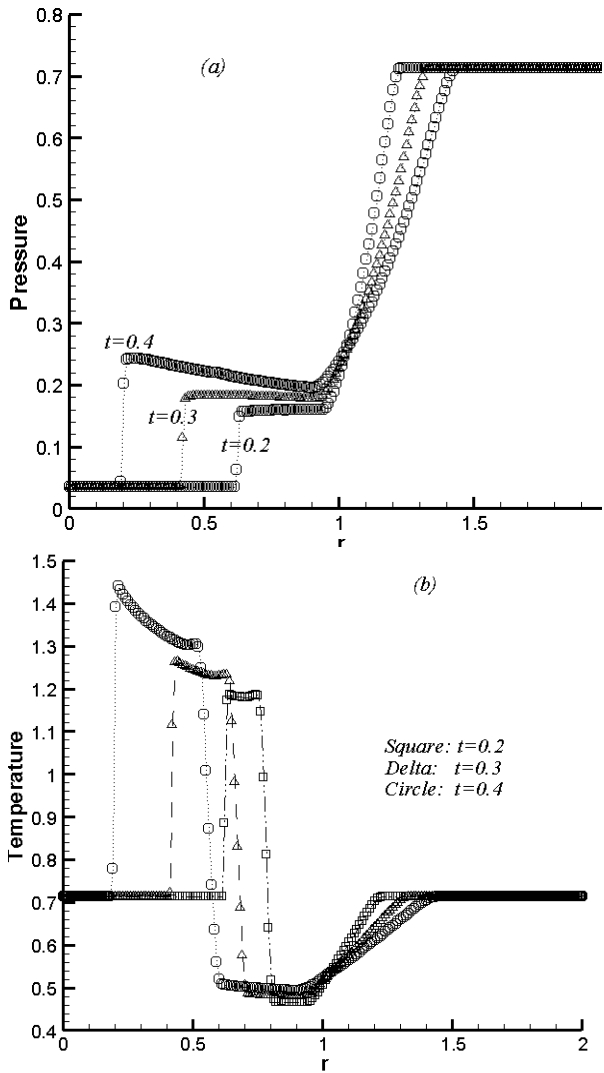

similar trend of variations as those of previous works even the present simulation of radially symmetric flows is advantageous over the method used by earlier researchers, because they are using schemes with different degree of artificial viscosity which may affect the physical situation and may change, up to certain extent, the fundamental principles of conservation.

With the most of other methods used in comparison, the shock and contact discontinuity are smeared and the smearing of contact discontinuity is more dramatic.

In the graphs we get steeper gradients than the most previous works will show. This leads to a better position of the shock wave and contact surface, and at the same time it indicates less dissipation.

In all the methods used earlier the shock converges at the axis of symmetry at a later or slightly earlier time than in our case. The time of convergence with artificial viscosity for others it is given as $t_c = 0.66$ (Payne), $t_c = 0.57$ (Abarbanel and Goldberg) and $t_c = 0.5826$ (Leutioff $et$ $al.$). As a criterion for the convergence-time the change of sign of the velocity in the axis point was applied.

## VI. CONCLUSION

The equation of gas dynamics are solved using a TVD scheme and MUSCL [1] approach based on Roe's approximate Riemann solver [2] is employed, along with the super-bee limiter [3] which keeps the shock waves and contact discontinuity perfectly sharp. The results obtained above clearly demonstrate the advantage of the scheme compared with the methods used by other authors. It seems to be a powerful method for computing flows with radialy symmetry and gives very good description of discontinuities during the simulations and the behavior of the solution in the focusing stage.

Fig. 3 Pressure (a) and Temperature (b) profiles at time intervals of 0.1 for initial pressure and density ratio 20.

This problem was solved by Payne [4] using Lax-difference scheme of first order and there is an implicit term of artificial viscosity in this scheme by approximation Abarbanel and Goldberg [6] have used the modified Lax-Wendroff scheme and an additional term is coming in their scheme which plays the role of artificial viscosity and takes the values between 0.2 and 1.0. Lapidus [5] has used the Cartesian method with Lax-Wendroff scheme, whereas Leutioff et al. [8] used the Cartesian method with Rusanov's scheme [13]. The artificial viscosity used in Leutioff et al. [8] calculations is about 0.04 and is hundred times smaller than the artificial viscosity used by Lapidus [5] which is 4.0. Though the distributions of pressure, Temperature and density in our case give the almost

### REFERENCES

[1] B. Van Leer, "Towards the ultimate conservative difference scheme. V.: A second-order sequel to Godunov's method ", *J. Comput. Phys.*, vol. 32, 1979, pp. 101-136.

[2] P. L. ROE, "Approximate Riemann Solvers, Parameter Vectors, and Difference Schemes", *J. Comput. Phys.*, vol. 43, 1981, pp. 357-372.

[3] P. L. ROE, "Characteristic-Based Schemes For The Euler Equations", *Annu. Rev. Fluid Mech.*, vol. 18, 1986, pp. 337-365.

[4] R. B. Payne, "A numerical method for a converging cylindrical shock", *J. Fluid Mech.*, vol. 2, 1957, pp. 185-200.

[5] A. Lapidus, "Computation of radially symmetric shocked flows", *J. Comput. Phys.*, vol. 8, 1971, pp. 106.

[6] S. Abarbanel and M. Goldberg, "Numerical solution of quasi-conservative hyperbolic systems of the cylindrical shock problem" *J. Comput. Phys.*, vol. 10, 1972, pp. 1.

[7] G. A. Sod, "A numerical study of a converging cylindrical shock", *J. Fluid Mech.*, vol. 83, 1977, pp. 785-794.

[8] D. Leutloff and K. G. Roesner, "Numerical Solution Of Converging Shock Problem", *Computers & Fluids*, vol. 16, No. 2, 1988, pp. 175-182.

[9] N. Satofuka, "A study of difference methods for radially symmetric shocked flows", *Lecture Notes in Physics*, (Edited by R. D. Richtmyer), 1975, Springer, Berlin.

[10] M. Ben-Artzi and J. Falcovitz, " An upwind second-order scheme for compressible duct flows ", SIAM J. Sci. Statist. Comput., vol. 7, No. 3, 1986, pp. 744-768.

[11] H. Matsuo, Y. Ohya and K. Fujiwara, "Numerical Simulation of Cylindrically Converging Shock Waves", *J. Comput. Phys.*, vol 75, 1988, pp. 384-399.

[12] J. Glimm, "Solution in the large for nonlinear hyperbolic system of equations", *Comm. Pure Appl. Math.*, vol. 18, 1965, pp. 697.

[13] V. V. Rusanov, "The calculation of non-stationary shock waves with obstacles", *Zh. Vychsl. mat. Fiz.*, vol. l, 1961.

[14] G. Guderley, "Starke kugelige und zylindrische Verdichtungsst6Be in der Nahe des Kugelmittelpunktes bzw. der Zylinderachse", *Luftfahrtforschung*, vol. 19, 1942, pp. 302.

[15] D. S. Butler, "Converging spherical and cylindrical shocks", Armaments Research Establishment, Rep. No. 54/55, 1954.

[16] Y. B. Zeldovich and Y. P. Raizer, *Physics of Shock Waves and High Temperature Hydrodynamic Phenomena—ll*, Academic Press, New York, 1967.

[17] G. B. Whitham, *Linear and Nonlinear Waves*, Wiley, New York, 1974.

[18] R. B. Lazarus and R. D. Richtmyer, Report LA-6823-MS, Los Alamos Sci. Lab., 1977.

[19] M. Van Dyke and A. J. Guttmann, " The converging shock wave from a spherical or cylindrical piston ", *J. Fluid Mech.*, vol. 120, 1982, pp. 451.

[20] P. Hafner, "Strong Convergent Shock Waves Near the Center of Convergence:A Power Series Solution", *SIAM J. Appl. Math.*, vol. 48, 1988, pp.1244.

[21] V. D. Sharma and CH. Radha, "Similarity Solutions For Converging Shocks In a Relaxing Gas", *Int. J. Engng Sci.*, vol. 33, No. 4, 1995, pp. 535-553.

[22] L.E. Bilbao and J. Gratton, "Spherical and cylindrical convergent shocks", *IL NUOVO CIMENTO*, vol. 18 D, No. 9, September 1996, pp. 1041-1060.

[23] R. Samtaney and D. I. Pullin, "On initial-value and self-similar solutions of the compressible Euler equations", *Phys. Fluids*, vol. 8 (10), October 1996, pp. 2650-2655.

[24] J. C. Hidalgoa and S. Mendozab, "Self-similar imploding relativistic shock waves", *Phys. Fluids*, vol. 17, (096101) 2005, pp. 1-8.

[25] J. Meyer-Ter-Vehn and C. Schalk "Self-similar compression waves in gas dynamics", *Z. Naturforsch*, vol. 37a, 1982, pp. 955-969.

[26] R. W. Perry and A. Kantrowitz, "The production and stability of converging shock waves", *J. appl. Phys.*, vol. 22, 1951, pp. 878.

[27] J. H. T. Wu, R. A. Neemeh, P. P. Ostrowski and M. N. Elabdin, "Production of converging cylindrical shock waves by finite element conical contractions", *Shock tube and shock wave research*, (Edited by Bayne Ahlborn, Abraham Hertzberg and David Russel), Seattle, 1977.

[28] J. H. T. Wu, R. A. Neemeh and P. P. Ostrowski, "Experimental studies of the production of converging cylindrical shock waves", *AIAA*, vol. 18, 1980.

[29] K. Takayama, H. Kleine, and H. Grönig, "An experimental investigation of the stability of converging cylindrical shock waves in air", *Exp. Fluids*, vol. 5, 1987, pp. 315.

[30] M. Watanabe, O. Onodera, K. Takayama, "Shock wave focusing in a vertical annular shock tube", *Shock waves at Marseille IV: Shock structure and kinematics*,1995.

[31] V. Eliasson, N. Tillmark, A. J. Szeri, and N. Apazidis, "Light emission during shock wave focusing in air and argon", *Phys. Fluids*, vol. 19, 2007.

[32] V. Eliasson, W. D. Henshaw, and D. Appelo, "On cylindrically converging shock waves shaped by obstacles", Physica D, 2008, pp. 1-7.

[33] M. Shabouei, R. Ebrahimi, and K. Mazaheri Body, "Numerical Simulation of Converging Spherical Shock Problem", In *Ninth International Congress of Fluid Dynamics & Propulsion (ICFDP 9)*, Alexandria, Egypt 2008.

[34] M. Komeili, M. Mirzaei, and M. Shabouei. "Performance of 1-D and 2-D Lattice Boltzmann (LB) in Solution of the Shock Tube Problem." In *International Conference on Fascinating Advancement in Mechanical Engineering (FAME2008)*, India, 2008, arXiv:1602.02675.